\documentclass{iau}
\usepackage{float}
\usepackage{graphicx}

\def\be{\begin{equation}}
\def\ee{\end{equation}}
\def\bea{\begin{eqnarray}}
\def\eea{\end{eqnarray}}

\def\dm{\Delta {\rm M}}
\def\ms{\ifmmode {\rm M}_{\odot}{ ~ } \else ${\rm M}_{\odot}{ ~ }$ { }\fi}

\def\mdot{\ifmmode \dot M \else $\dot M$\fi}    
\def\mxd{\ifmmode \dot {M}_{x} \else $\dot {M}_{x}$\fi}
\def\med{\ifmmode \dot {M}_{Edd} \else $\dot {M}_{Edd}$\fi}
\def\bff{\ifmmode B_{{\rm f}} \else $B_{{\rm f}}$\fi}

\def\apj{\ifmmode ApJ \else ApJ \fi}    
\def\apjl{\ifmmode  ApJ \else ApJ \fi}    %
\def\apjs{\ifmmode  ApJS \else ApJS \fi}
\def\aap{\ifmmode A\&A \else A\&A\fi}
\def\aaps{\ifmmode A\&AS \else A\&AS\fi}    %
\def\mnras{\ifmmode MNRAS \else MNRAS \fi}    %
\def\nat{\ifmmode Nature \else Nature \fi}
\def\prl{\ifmmode Phys. Rev. Lett. \else Phys. Rev. Lett.\fi}
\def\prd{\ifmmode Phys. Rev. D. \else Phys. Rev. D.\fi}
\def\pasp{\ifmmode  PASP \else PASP \fi}

\def\ms{\ifmmode {\rm M}_{\odot} \else ${\rm M}_{\odot}$\fi}    
\def\na{\ifmmode \nu_{A} \else $\nu_{A}$\fi}    
\def\nk{\ifmmode \nu_{K} \else $\nu_{K}$\fi}    
\def\ns{\ifmmode \nu_{{\rm s}} \else $\nu_{{\rm s}}$\fi}
\def\no{\ifmmode \nu_{1} \else $\nu_{1}$\fi}    
\def\nt{\ifmmode \nu_{2} \else $\nu_{2}$\fi}    
\def\ntk{\ifmmode \nu_{2k} \else $\nu_{2k}$\fi}    
\def\dnmax{\ifmmode \Delta \nu_{max} \else $\Delta \nu_{2max}$\fi}
\def\ntmax{\ifmmode \nu_{2max} \else $\nu_{2max}$\fi}    
\def\nomax{\ifmmode \nu_{1max} \else $\nu_{1max}$\fi}    
\def\nh{\ifmmode \nu_{\rm HBO} \else $\nu_{\rm HBO}$\fi}    
\def\nqpo{\ifmmode \nu_{QPO} \else $\nu_{QPO}$\fi}    
\def\nz{\ifmmode \nu_{o} \else $\nu_{o}$\fi}    
\def\nht{\ifmmode \nu_{H2} \else $\nu_{H2}$\fi}    
\def\ns{\ifmmode \nu_{s} \else $\nu_{s}$\fi}    
\def\nb{\ifmmode \nu_{{\rm burst}} \else $\nu_{{\rm burst}}$\fi}
\def\nkm{\ifmmode \nu_{km} \else $\nu_{km}$\fi}    
\def\ka{\ifmmode \kappa \else \kappa\fi}    
\def\dn{\ifmmode \Delta\nu \else \Delta\nu\fi}
\def\dm{\ifmmode \Delta{}M \else \Delta{}M\fi}
\def\mdotsix{\ifmmode\dot{M}_{16} \else \dot{M}_{16}\fi}
\def\ps{\ifmmode P_{spin} \else P_{spin} \fi}

\def\sax{\ifmmode SAX J 1808.4-3658 \else SAX J 1808.4-3658\fi}

 \def\pspin{\ifmmode P_{s} \else $P_{s}$\fi}

\def\rhof{\ifmmode \rho_{5} \else \rho_{5}\fi}
\def\rhos{\ifmmode \rho_{6} \else \rho_{6}\fi}
\def\mdotcr{\ifmmode \dot{M}_{cr} \else  \dot{M}_{cr}\fi}
\def\tohm{\ifmmode t_{ohmic} \else  t_{ohmic} \fi}

\def\tdif{\ifmmode t_{diff} \else  t_{diff} \fi}
\def\tacc{\ifmmode t_{accr} \else  t_{accr} \fi}

\title[Minimum accretion rate for millisecond pulsar formation] 
{Minimum accretion rate for millisecond pulsar formation in binary system}

\author[Yuanyue Pan, Chengmin Zhang and Na Wang]   
{Yuanyue Pan$^{1,2}$, Chengmin Zhang$^{1}$
\and Na Wang$^{2}$ }

\affiliation{$^{1}$National Astronomical
Observatories, Chinese Academy of Sciences, Beijing 100012,
China\\Email: {\tt panyuanyue@xao.ac.cn} \\
$^{2}$Xinjiang  Astronomical Observatories, Chinese Academy of
Sciences, Xinjiang 830011, China\\ }

\pubyear{2012}
\volume{290}  
\jname{Feeding compact objects: Accretion on all scales}
\editors{C.M. Zhang, T. Belloni, M. M\'endez \& S.N. Zhang, eds.}
\begin{document}

\maketitle

\begin{abstract}
186 binary pulsars are shown in the magnetic field versus spin period (B-P)
diagram, and their relations to the millisecond pulsars can be clearly seen.
We declaim a minimum accretion rate for the millisecond pulsar formation both from
the observation and theory. If the accretion rate is lower than the minimum accretion
rate, the pulsar in binary system will not become a millisecond pulsar after the
evolution.
\end{abstract}

\firstsection 

\section{Analyzing the binary pulsars in B-P diagram}
Until now, it has been found 186 binary radio pulsar systems, including 136 millisecond
pulsars with the spin period less than 20 milliseconds.
Much progress has been achieved in
understanding the formations and evolutions of the binary pulsars
(\cite[Bhattacharya \& van den Heuvel 1991]{bha91}; \cite[Stairs 2004]{sta04};
\cite[Tauris 2012]{tau12}). In a binary system, with the accreting matter of
$\sim 0.1-0.2 M_{\odot}$ from the companion, a neutron star can be spin
up to several milliseconds, while its magnetic field will be changed from
$\sim 10^{12} G$ to $\sim 10^{8-9} G$ (
\cite[Zhang \& Kojima 2006]{zhang06}; \cite[Wang, Zhang \& Zhao 2011]{wang11}).
There exist some possible cases that the single millisecond pulsar can be
formed due to the evaporation of companion by the millisecond pulsar
radiation (\cite[Kluzniak et al. 1988]{klu88}). Direct supporting of the millisecond
pulsar spin-up model has been found, for instance, the accreting X-ray millisecond pulsar
SAX J1808.4-3658 with $B\sim10^{8} G$ and $P=2.49  ms$ is discovered in low mass
X-ray binary system (\cite[Wijnands \& van der Klis 1998]{wij98}), and double
pulsars PSR0737-3039A/B are found with one millisecond pulsar and a normal
pulsar (\cite[van den Heuvel 2004]{heu04}; \cite[Lyne 2004]{lyn04}).

\begin{figure}[H]
\label{bpsr186}
\centering
\includegraphics[width=2in]{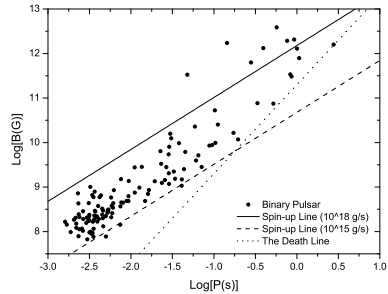}
\caption{The magnetic field versus spin period diagram for 186 binary
pulsars (data from ATNF pulsar catalogue, see Manchester et al. 2005).
The solid, dash and dot lines are the spin-up line with the accretion rates
$10^{18} g/s$, $10^{15} g/s$ and the death line, respectively.}
\end{figure}

The spin-up line in Fig.1 is also called the equilibrium spin period
line (\cite[Bhattacharya \& van den Heuvel 1991]{bha91}), 
which expresses the relationship between the magnetic field and the spin
period, when a neutron star is spun up to the Kepler orbital period at the magnetosphere
radius. 
The equation of the spin-up line can be written as: 
$P_{eq}=2.4 (ms)B_9^{6/7}(M/M_{\odot})^{-5/7}(\dot{M}/\dot{M}{_{Edd}})^{-3/7}R_6^{16/7}$,
where $P_{eq}$ is the "equilibrium" spin period, $B_9$ is the magnetic field in
units of $10^9G$, $\dot{M}$ is the accretion rate, $\dot{M}_{Edd}$ is the maximum
possible "Eddington-limit" accretion rate ($10^{18} g/s$), $R_6$ is the stellar
radius in units of $10^6$cm and $M$ is the neutron star mass. Let the accretion rate be
$10^{18} g/s$ and $10^{15} g/s$ in the spin-up line equation, 
there will be two spin-up lines as shown
in Fig.\ref{bpsr186}. With the 186 binary pulsars given in the figure, it can be seen that
almost all of the binary pulsars are above the spin-up line with the accretion rate
$10^{15} g/s$. However, there are three exceptions: J0514-4002A, J1801-3210 and
J1518+4904. These three binary pulsars lie below the spin-up line with
$\dot{M}=10^{15} g/s$, which correspond to the inferred accretion rates
$0.6\times 10^{15}g/s, 0.8\times 10^{15}g/s$ and $0.9\times 10^{15}g/s$,
respectively. Thus from the distribution of the binary pulsars in Fig.1,
it can be suggested that $10^{15} g/s$ is a critical minimum accretion rate
for the millisecond pulsar formation.
The real ages of millisecond pulsars have not yet been satisfactorily
determined, although their characteristic ages can be as long as
the Hubble time. 
During the millisecond pulsar spin-up evolution, the maximum critical accreting mass
for a millisecond pulsar to reach the spin period of several milliseconds is about
$\Delta M_{cr} \sim 0.1-0.2 M_{\odot}$ (\cite[Zhang, Wang \& Zhao et al.
2011]{zhang11}; \cite[Wang, Zhang \& Zhao et al. 2011]{wang11}),
and the time of spinning-up should be less than the Hubble age of about
$t_{H}=10^{10} yrs$. Thus, a critical mass accretion rate for a millisecond pulsar formation
can be evaluated by $\dot{M}_{cr} = \Delta M_{cr} / t_{H} \simeq 10^{15} g/s\;.$
If the accretion rate is much lower than this critical value, the
neutron star should have no chance to accrete sufficient mass to be
spun-up to a spin period of several milliseconds. 

\section{Conclusion}

The 186 binary pulsars in the B-P diagram show us that almost all binary
pulsars lie above the minimum spin-up line with $\dot{M}=10^{15} g/s$. In
theory, it can also be deduced that with the maximum accretion mass within
Hubble time in binary system, the accretion rate will be $\dot{M}\geq10^{15} g/s$.
If the accretion rate is lower than that value, a pulsar in the binary system will
have no chance to become a millisecond one even with the maximum accretion mass during the
evolution. So, it can be conclude that $\dot{M}=10^{15} g/s$ is the minimum
rate for the millisecond pulsar formation in the binary system.
%

This work is supported by National
Basic Research Program of China (973 Program 2009CB824800), China Ministry of Science and
Technology under State Key Development Program for Basic Research
(2012CB821800), NSFC10773017,  NSFC11173034, and Knowledge Innovation Program of CAS KJCX2-
YW-T09.

\end{document}